\documentclass[%
 reprint,
 amsmath,amssymb,
 aps,
prb,
]{revtex4-2}

\usepackage{placeins}
\usepackage{graphicx}
\usepackage{dcolumn}
\usepackage{comment}
\usepackage{xcolor}
\usepackage{bm}

\usepackage{verbatim}

\usepackage{fancyhdr}
\usepackage{datetime}
\newdateformat{monthyeardate}{\shortmonthname[\THEMONTH] \THEYEAR}

\begin{document}

\preprint{APS/123-QED}

\title{Berry Curvature Dipole-Induced Chiral Terahertz Gain and Lasing \textcolor{black}{Threshold} in Bulk Tellurium
}

\author{Mounes Eslami$^1$}
 \email{mounes.eslami@gmail.com}
\author{Amin Hakimi$^1$}
 \email{amin.hakimi@uci.edu}
\author{Luis A. Jauregui$^2$}
 \email{lajaure1@uci.edu}
\author{Filippo Capolino$^1$}
 \email{f.capolino@uci.edu}

\affiliation{$^1$Department of Electrical Engineering and Computer Science, University of California, Irvine, CA 92697, USA\\
$^2$Department of Physics and Astronomy, University of California, Irvine, CA 92697, USA}

%


\begin{abstract}
We investigate the use of Berry curvature dipole in $n$-doped Tellurium as a mechanism for achieving terahertz amplification and lasing by applying a DC electric field. When the electrical bias and wave vector are aligned along the trigonal $c$-axis, the right-handed circularly polarized mode experiences amplification at relatively low bias, while the left-handed mode is attenuated. Furthermore, when the electrical bias and wave vector are orthogonal to the $c$-axis, the structure supports elliptically polarized eigenmodes that also exhibit gain under suitable bias conditions, where the degree of ellipticity is tunable by the applied bias. We also investigate lasing conditions for a Fabry-Perot cavity incorporating biased Te as an active medium. Due to the resonance in the dielectric permittivity of Tellurium, there are discrete lasing intervals. Our results show that bulk chiral Tellurium could be used as an electrically tunable, polarization-selective gain medium for micrometer-scale terahertz lasers, with lasing achievable at bias fields below the material's breakdown threshold, paving the way towards new terahertz devices.

\end{abstract}

\maketitle
\section{Introduction}
Exploring quantum materials with novel topological phases and probing their symmetry properties remains a central goal of condensed matter physics. Topological behavior in materials is often governed by the Berry curvature, a fundamental geometric property of electronic bands that gives rise to phenomena such as the quantum Hall effect \cite{xiao2010berry, nagaosa2010anomalous} in systems that break time-reversal symmetry. The Berry curvature captures the topological structure of electronic bands \cite{Thouless1982, Berry1984, xiao2010berry, bansil2016colloquium, narang2021topology}. An imbalanced Berry curvature distribution can serve as an indicator for probing the geometric deformation in the band structure, leading to the emergence of the Berry curvature dipole (BCD), defined as the first moment of the Berry curvature in momentum space. In non-centrosymmetric materials, the BCD plays a key role in driving nonlinear quantum phenomena such as the nonlinear Hall effect (NLHE) \cite{sodemann2015quantum, du2018band, ma2019observation}, the anomalous planar Hall effect (APHE) \cite{wang2023field, battilomo2021anomalous}, the quantized circular photogalvanic effect \cite{de2017quantized, rees2020helicity}, and the quantum shift current \cite{cook2017design, zhang2019enhanced, osterhoudt2019colossal, akamatsu2021van, holder2020consequences}, even in time-reversal invariant systems. This nonlinear effect has garnered significant attention, particularly in connection with the growing interest in nonlinear optical and transport phenomena in quantum materials, as it has been used in the rectification of alternating currents \cite{kumar2021room, hu2023terahertz, lu2024nonlinear, sinha2022berry}.

Recent studies have revealed that the BCD not only governs electronic transport properties but also modifies the optical response of materials. For example, a large BCD leads to giant second harmonic generation \cite{fu2023berry}. Moreover, electro-optic (EO) effects in the materials exhibiting BCD can induce nonreciprocal optical gain, which originates from the non-Hermitian component of the EO response \cite{rappoport2023engineering, hakimi2024chiral, lannebere2025symmetry}.
\textcolor{black}{Unlike conventional stimulated-emission-based amplification, which relies on population inversion, the mechanism here emerges from nontrivial intraband Bloch-electron dynamics in the presence of static fields \cite{roldan2025optical} that interact with the optical wave. The optical wave would drive the electrons to emit additional radiation in phase with it, leading to amplification, in a way conceptually similar to what occurs in a free-electron laser \cite{o2001free} or in an electron cyclotron maser (or gyrotron) \cite{chu2004electronCyclo}.}

Chiral materials with large BCD result in polarization-dependent coupling between the electronic states and the incident light \cite{morgado2024non, liu2024anomalous}. This concept has been further investigated for chiral lasing in low-symmetry two-dimensional (2D) materials with large BCD, such as twisted bilayer graphene \cite{hakimi2024chiral}. Weyl semimetals, topological states with linear energy dispersion crossing at Weyl nodes, which act as momentum-space sources of Berry curvature and BCD \cite{zhang2018berry} have also been identified as promising platforms. There are also other effects induced in nonmagnetic materials by BCD, including the kinetic Faraday effect, a change in rotatory power caused by an electrical current \cite{vorob1979optical, tsirkin2018gyrotropic}.

Recent theoretical studies have shown BCD-induced amplification in Tellurium (Te) using both bulk and edge modes \cite{morgado2024non, prudencio2025topological}. Trigonal Te is a chiral semiconductor \cite{asendorf1957space, tsirkin2018gyrotropic, niu2023tunable}, which exhibits a large intrinsic BCD and has demonstrated a pronounced NLHE in both bulk and 2D forms \cite{joseph2024chirality, cheng2024giant}. This makes Te a promising candidate for applications in frequency-doubling and rectifying devices, based on the general mechanism of chiral Bloch‑electron rectification in inversion‑breaking semiconductors \cite{isobe2020high}. Recent theoretical studies predict that $n$-doped Tellurium, due to its large values of BCD \cite{morgado2024non}, can achieve optical field amplification with lower static electric field biases compared to $p$-doped Te. In this work, we build on the conductivity calculations developed in \cite{morgado2024non} for three-dimensional (3D) Tellurium exhibiting a non-Hermitian linear EO effect to investigate how different configurations of static electric field and wave propagation direction can be exploited to realize tunable optical gain in $n$-doped Tellurium. Our results demonstrate that Tellurium can serve as an active medium for micrometer-scale terahertz (THz) lasers and cavities by exploiting BCD, offering a novel platform for chiral light amplification and generation in bulk, nonmagnetic semiconductors. \textcolor{black}{This analysis focuses on threshold conditions and modal selectivity for BCD-induced gain in Te, rather than on a full laser dynamics model. Gain saturation, mode competition, and nonlinear temporal dynamics are beyond the scope of this study. Nevertheless, threshold analysis and cavity-mode selection provide essential guidance for assessing the feasibility and operating regimes of electrically driven chiral THz sources.}

\textcolor{black}{
Conventional THz wave generation has been based on methods that can be broadly categorized into electronic, optical, and hybrid techniques, depending on the desired frequency range (0.1–10 THz), power, coherence, and application \cite{zhang2021intense}. Semiconductor and quantum devices enable \textit{direct} THz generation, for example, quantum cascade lasers (QCLs), which use intersubband gain in repeated quantum wells such as GaAs/AlGaAs or InGaAs/AlInAs \cite{gao2023recent}. In contrast, most electronic and optical methods rely on \textit{indirect} mechanisms such as frequency mixing or difference-frequency generation. 
Here, we focus on a totally different way for {\em direct} THz generation using a DC pump applied to BCD materials. Compared to QCLs, this Berry-curvature-driven mechanism may offer some advantages to be explored. QCLs face thermal limitations, typically requiring cryogenic or sub-room-temperature operation \cite{kainz2018barrier} and primarily operating within the $\sim$1–5 THz range \cite{shahili2024continuous}. In contrast, our proposed approach can lead to broader THz bandwidths and more flexible frequency operation because the Berry-curvature response is not tied to a narrow intersubband transition frequency \cite{seifert2021frequency}. While BCD materials can also be thermally sensitive, materials with robust band structures can retain a large BCD even at room temperature \cite{nishijima2023ferroic}. Additionally, placing the gain medium inside a cavity appears to permit the use of smaller cavities than the millimeter-scale ones typically used in QCLs \cite{young2009wavelength}, facilitating compact and miniaturized THz sources.}

This paper is organized as follows. Section \ref{Sec:amplification} introduces three configurations of tellurium–light interaction, each characterized by a specific bias and propagation direction. For each case, the dispersion relation is analyzed to determine the polarization eigenstates and wavenumbers. In Section \ref{Sec:lasing}, the lasing condition is solved for bulk Tellurium modeled as a gain medium in a simple Fabry–Perot laser, examining how the cavity length and mirror reflectivity influence its performance. Finally, Section \ref{Sec:conclusions} presents the conclusions and summarizes the key findings.

\section{Chiral Amplification}
\label{Sec:amplification}
According to \cite{morgado2024non}, when a static electric field bias $\mathbf{E_0}$ is applied to a bulk material with a BCD tensor $\underline{\mathbf{D}}_{\rm B}$, the general expression for EO conductivity using Boltzmann semiclassical approximation is given by 
\begin{align}
    \underline{\boldsymbol{\sigma}}_{\mathrm{EO}}(\omega)
    &= \underline{\boldsymbol{\sigma}}_{\mathrm{EO}}^{\mathrm{G}} + 
    \underline{\boldsymbol{\sigma}}_{\mathrm{EO}}^{\mathrm{NH}}(\omega)\notag \\ 
    &= - \frac{\tau e^3}{\hbar^2} \left( \mathbf E_0 \cdot \underline{\mathbf{D}}_{\rm B} \right)\times \underline{\mathbf{1}}
    +
    \frac{\tau e^3}{\hbar^2} \frac{1}{1 - i \omega \tau} \left(\mathbf E_0 \times\underline{\mathbf{D}}_{\rm B}^{{\rm T}}\right),
    \label{eq:EOGeneralConductivity}
\end{align}
where $\tau$ denotes the scattering relaxation time, $\omega$ is the angular frequency of the applied (optical) field when probing the dynamic response, and $e$ is the elementary charge. The symbol $\underline{\mathbf{1}}$ represents the identity tensor, and ${\rm T}$ indicates the transpose operator. We adopt the time-harmonic convention $e^{-i\omega t}$ throughout this paper.

\begin{figure*}[bt!]
	\centering
	\includegraphics[scale=0.4]{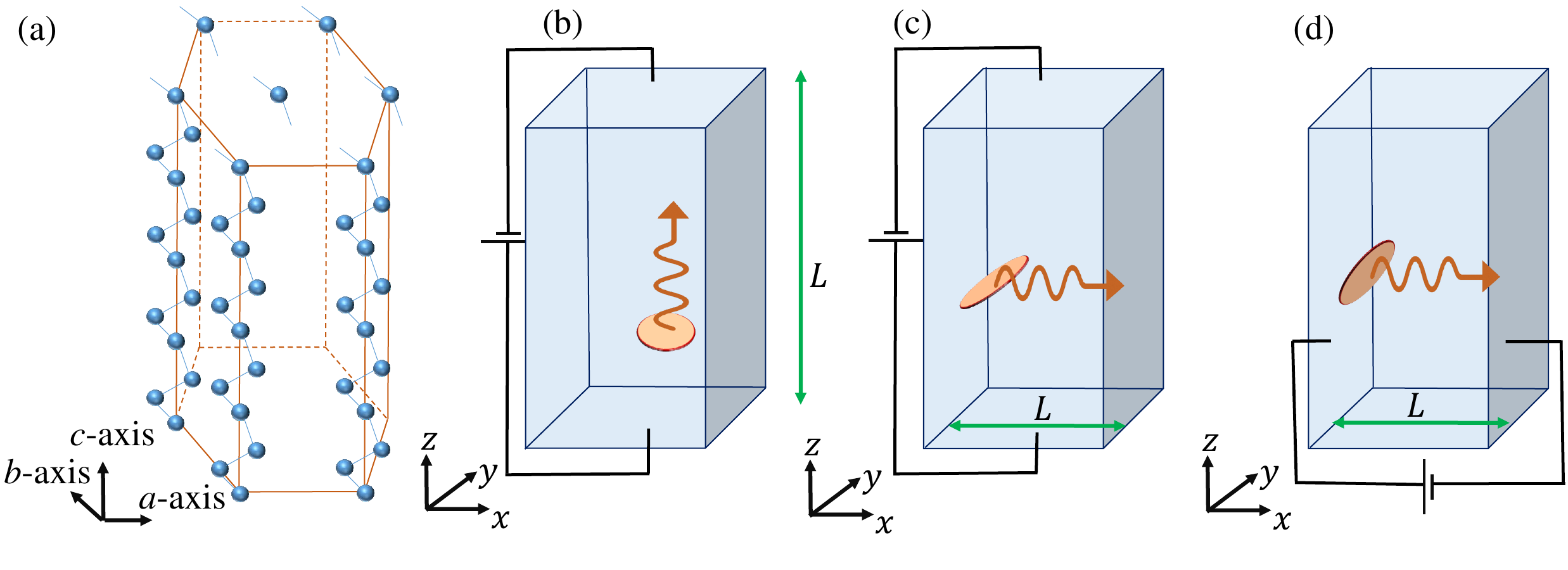} 
	\caption{Schematic illustration of DC-biased tellurium interacting with optical fields. (a) Molecular packing of right-handed Tellurium showing the crystallographic $a$-, $b$- and $c$-axes relative to the laboratory-frame $x$, $y$, $z$ coordinate system \cite{asendorf1957space}. (b) Case I: Both the optical wave propagation direction and the applied bias are aligned along the $z$ direction. (c) Case II: The bias is applied along $z$, while the optical wave propagates along $x$. (d) Case III: both the optical wave propagation direction and the applied bias are aligned along the $x$ direction. In all cases, $L$ is the length of the cavity along which the lasing occurs.} 
	\label{fig:generalViewofCases}
\end{figure*}

The total conductivity of Tellurium  $\underline{\boldsymbol{\sigma}}(\omega) = \underline{\boldsymbol{\sigma}}_{\mathrm{D}}(\omega) + \underline{\boldsymbol{\sigma}}_{\mathrm{EO}}(\omega)$, accounts also of the standard Drude's model 
\begin{equation}
\underline{\boldsymbol{\sigma}}_{\mathrm{D}}(\omega) = \frac{\varepsilon_0 \omega_{\mathrm{p},l}^2}{\Gamma - i\omega} \underline{\mathbf{1}},
    \label{eq:DrudeConductivity}
\end{equation}
where $\Gamma = 1/\tau$ is the collision frequency, and $\omega_{\mathrm{p},l}$ denotes the plasma frequency. Here, $l = \mathrm{\perp}, \mathrm{||}$ denote the perpendicular and parallel directions with respect to the trigonal axis ($c$-axis), respectively. The $x-$ and $z-$axes correspond to the crystallographic $a$-axis and $c$-axis, respectively, as illustrated in Fig. \ref{fig:generalViewofCases}(a).

The total effective permittivity tensor is 
\begin{equation}
    \frac{\underline{\boldsymbol{\varepsilon}}(\omega)}{\varepsilon_0} = \underline{\boldsymbol{\varepsilon}}_\mathrm{d}(\omega) + i \frac{\underline{\boldsymbol{\sigma}}(\omega)}{\varepsilon_0 \omega},
    \label{eq:GeneralPermittivity}
\end{equation}
where $\underline{\boldsymbol{\varepsilon}}_\mathrm{d}(\omega)$ denotes the dielectric response of the bulk material.
 \textcolor{black}{The first term in Eq. (\ref{eq:EOGeneralConductivity}) represents the gyrotropic part of the conductivity and the associated effective permittivity \cite{Note_1_Gyro} is given by $\underline{\boldsymbol{\varepsilon}}_{\rm EO}^{\rm G} = i \underline{\boldsymbol{\sigma}}_{\mathrm{EO}}^{\mathrm{G}}/\omega$. It is Hermitian, i.e., not associated with energy exchange between the optical wave and the material. Nevertheless, the gyrotropic contribution is essential for accurately describing wave propagation in the material. The gyrotropic part of the electric flux density is given by $\mathbf{D}^{\rm G}=\underline{\boldsymbol{\varepsilon}}_{\rm EO}^{\rm G} \cdot \mathbf{E} = i \mathbf{E} \times \mathbf{g}$ where $\mathbf{E}$ is the optical electric field. Here,  $\mathbf{g}= \varepsilon_0  \boldsymbol{\omega}_{\mathrm{c}}/\omega$ is the gyration vector \cite{landau2013electrodynamics} that is proportional to an effective cyclotron-like (or gyrotron-like) angular frequency vector 
$\boldsymbol{\omega}_{\mathrm{c}} = \frac{\tau e^3}{\varepsilon_0 \hbar^2}\left( \mathbf E_0 \cdot \underline{\mathbf{D}}_{\rm B} \right)$.}

For Te, the BCD tensor is given by the dyad $\underline{\mathbf{D}}_{\rm B} = D_{\rm B} \hat{\mathbf{x}} \hat{\mathbf{x}} + D_{\rm B} \hat{\mathbf{y}} \hat{\mathbf{y}} -2D_{\rm B} \hat{\mathbf{z}} \hat{\mathbf{z}}$ \cite{tsirkin2018gyrotropic}. Assuming a static electric field of the form $\mathbf{E_0} = E_{0, x} \hat{\mathbf{x}} + E_{0, y} \hat{\mathbf{y}} + E_{0, z} \hat{\mathbf{z}}$, in Cartesian coordinates the matrix form of the relative permittivity becomes 
\begin{equation}
\frac{\underline{\boldsymbol{\varepsilon}}}{\varepsilon_0} = \begin{bmatrix}
\varepsilon_{\mathrm{\perp}} &  \varepsilon_{\mathrm{EO}, xy} &  \varepsilon_{\mathrm{EO}, xz}\\
\varepsilon_{\mathrm{EO}, yx} & \varepsilon_{\mathrm{\perp}} &  \varepsilon_{\mathrm{EO}, yz} \\
\varepsilon_{\mathrm{EO}, zx} & \varepsilon_{\mathrm{EO}, zy} & \varepsilon_{\mathrm{||}}
\label{eq:TelluriumGeneralPermittivity}
\end{bmatrix},
\end{equation}
where the off-diagonal electro-optic components are 
%
\begin{equation}
\begin{split}
&\varepsilon_{\mathrm{EO}, xy} =-\varepsilon_{\mathrm{EO}, yx}  = i \frac{\omega_{\mathrm{c}, z} }{\omega} \left(1 + \zeta/2 \right), \\
& \varepsilon_{\mathrm{EO}, xz} = -i\frac{ \omega_{\mathrm{c}, y} }{\omega} \left( 1 + 2\zeta \right), \\
& \varepsilon_{\mathrm{EO}, zx} = i\frac{ \omega_{\mathrm{c}, y} }{\omega} \left( 1 - \zeta \right), \\
& \varepsilon_{\mathrm{EO}, yz} = i\frac{ \omega_{\mathrm{c}, x} }{\omega} \left( 1 + 2\zeta \right), \\
& \varepsilon_{\mathrm{EO}, zy} = -i\frac{ \omega_{\mathrm{c}, x} }{\omega} \left( 1 - \zeta \right),
\end{split}
\label{eq: relativeEpsilonEO}
\end{equation}
with $\zeta = \Gamma/ (\Gamma - i \omega)$ and the effective cyclotron-like angular frequencies 
\begin{equation}
\omega_{\mathrm{c}, q} = \frac{\tau e^3}{\varepsilon_0 \hbar^2} D_{\rm B} E_{0, q}, \quad \omega_{\mathrm{c}, z} = -\frac{\tau e^3}{\varepsilon_0 \hbar^2} 2D_{\rm B} E_{0, z},
\label{eq:cyclotron_frequency}
\end{equation}
with $(q = x, y)$ are the component of the vector $\boldsymbol{\omega}_{\mathrm{c}}$ defined earlier. In Tellurium, the noncentrosymmetric structure induces off-diagonal permittivity elements in the presence of a DC bias field (as calculated in Eq. \eqref{eq: relativeEpsilonEO}). \textcolor{black}{The first term in Eq. \eqref{eq:EOGeneralConductivity} that yields the $\zeta$-independent contribution in Eq.\eqref{eq: relativeEpsilonEO}} represents the gyrotropic \cite{Note_1_Gyro,landau2013electrodynamics} EO response of Te, \textcolor{black}{with gyration vector $\mathbf{g}$ and associated effective cyclotron-like frequency providing the nonreciprocal effect. This nonreciprocity is fundamentally different from the magnetic-field–induced cyclotron effects in magneto-optical materials, as it is generated by electrostatic bias and not by an external magnetic field}. Furthermore, the gyrotropic EO effect does not contribute to either loss or gain because it is Hermitian. In contrast, the non-Hermitian part of the EO effect,  responsible for the $\zeta$ term in Eq. (\ref{eq: relativeEpsilonEO}),  contributes to polarization‑dependent gain or loss. Furthermore, the diagonal elements in Eq. (\ref{eq:TelluriumGeneralPermittivity}) are 
\begin{equation}
\varepsilon_{l} = \varepsilon_{\mathrm{d}, l} - \frac{\omega_{\mathrm{p}, l}^2}{\omega^2 + i\Gamma \omega}.
\label{eq:digonalelement}
\end{equation}
The material parameters for $n$-doped right-handed Tellurium are listed in Table \ref{tab:constantValues}, as computed in \cite{morgado2024non}. 

\begin{table}[h]
    \centering
    \caption{Material parameter values for $n$-doped Te.}
    \begin{tabular}{|c|c|}
        \hline
        Variable & Value \\ 
        \hline
        $\Gamma$ & $1.5625 \times 10^{12}\:\mathrm{rad/s}$ \\ 
        $n$ & $0.784 \times 10^{16}\:\mathrm{cm}^{-3}$ \\ 
        $D_{\rm B}$ & $-8.99 \times 10^{-3}$ \\ 
        $\omega_{\mathrm{p}, \perp}/(2\pi)$ & $1.857 \times 10^{12}\:\mathrm{Hz}$\\ 
        $\omega_{\mathrm{p}, \parallel}/(2\pi)$ & $2.345 \times 10^{12}\:\mathrm{Hz}$\\
        \hline
    \end{tabular}
    \label{tab:constantValues}
\end{table}

Additionally, the dielectric response of Tellurium in the frequency range from 0.1 THz to 30 THz, appearing in the first term of Eq. \eqref{eq:digonalelement}, using the Lorentz model \cite{palik1998handbook}, is 
\begin{gather}
\varepsilon_{\mathrm{d},\mathrm{\perp}} (\omega) = 23 - \frac{21.23 \times 10^{26}}{\omega^2 - i 5.654 \times 10^{11} \omega - 3.033 \times 10^{26}}, \label{eq:DielectricPermittivityLorentz1} \\
\varepsilon_{\mathrm{d}, \mathrm{||}} (\omega) = 36 - \frac{19.85 \times 10^{26}}{\omega^2 - i 6.597 \times 10^{11} \omega - 2.757 \times 10^{26}}.
\label{eq:DielectricPermittivityLorentz2}
\end{gather}
where $\omega$ is in units of radians per second. Resonances in dielectric response happen at approximately 2.64 THz and 2.76 THz for perpendicular and parallel polarizations, respectively, \textcolor{black}{originating from optical phonon resonance \cite{kittel2018introduction}.} \textcolor{black}{The term $\varepsilon_{\mathrm{d}, l}$  in Eq.~(\ref{eq:digonalelement}) is attributed solely to bound electrons and is therefore assumed to be independent of doping concentration.}  \textcolor{black}{Note that in \cite{morgado2024non}, the authors reported a single-frequency estimate of the dielectric permittivity from the experimental data in \cite{grosse1970absorption}. Here, because our analysis considers a broader frequency range (0.1–30 THz), we adopt the full Lorentzian model in Eqs.~\eqref{eq:DielectricPermittivityLorentz1} and \eqref{eq:DielectricPermittivityLorentz2}.}

Wave propagation in anisotropic materials is governed by the wave equation
$\nabla \times (\nabla \times \mathbf{E}) = \omega^2 \mu \underline{\boldsymbol{\varepsilon}}\cdot\mathbf{E}$ that, when assuming a plane wave with wavevector $\mathbf{k} = k_x \hat{\mathbf{x}} + k_y \hat{\mathbf{y}} + k_z \hat{\mathbf{z}}$, reduces to 
\begin{equation}
\mathbf{k} \times (\mathbf{k} \times \mathbf{E}) + \omega^2 \mu \underline{\boldsymbol{\varepsilon}}\cdot\mathbf{E} = 0.
\label{eq:WaveEquation_k}
\end{equation}
Using the permittivity matrix in Eq. \eqref{eq:TelluriumGeneralPermittivity}, the dispersion relation for the supported modes is obtained by  \cite{amnon1984optical}
%
\begin{widetext}
\begin{equation}
    \det
    \begin{bmatrix}
        k_y^2 + k_z^2 - k_0^2 \varepsilon_{\mathrm{\perp}} & -k_x k_y- k_0^2 \varepsilon_{\mathrm{EO}, xy} & -k_x k_z - k_0^2 \varepsilon_{\mathrm{EO}, xz} \\
        -k_y k_x - k_0^2 \varepsilon_{\mathrm{EO}, yx} & k_x^2 + k_z^2 - k_0^2 \varepsilon_{\mathrm{\perp}} & -k_y k_z - k_0^2 \varepsilon_{\mathrm{EO}, yz}\\
        -k_z k_x - k_0^2 \varepsilon_{\mathrm{EO}, zx} & -k_z k_y - k_0^2 \varepsilon_{\mathrm{EO}, zy} & k_x^2 + k_y^2- k_0^2 \varepsilon_{\mathrm{||}}
    \end{bmatrix}
    = 0,
    \label{eq:GeneralDispersion}
\end{equation}
\end{widetext}
where $k_0 = \omega \sqrt{\mu_0 \varepsilon_0}$ is the free-space wavenumber.
Solving Eq. (\ref{eq:GeneralDispersion}) yields the wavenumbers (eigenvalues), while the corresponding polarization states (eigenvectors) of the optical fields are obtained from the null space of the associated matrix.

To observe a chiral gain-dissipative response, one eigenmode typically undergoes amplification while the other experiences dissipation. Amplification occurs when the real and imaginary parts of the complex wavenumber, $\beta = \Re(k)$ and $\alpha = \Im(k)$, have opposite signs. However, to rigorously confirm amplification or attenuation, we also examine the energy transfer from the material to each eigenmode by evaluating the dissipated power density, which is given by \cite{Felsen1973radiation},
\begin{align}
p_{\mathrm{dis}} &= \frac{1}{2} \omega \mathbf{E}^* \cdot \underline{\boldsymbol{\varepsilon}}'' \cdot \mathbf{E},
\label{eq:GeneralDissipatedPower}
\end{align}
where $\underline{\boldsymbol{\varepsilon}}'' = \left(\underline{\boldsymbol{\varepsilon}} - \underline{\boldsymbol{\varepsilon}}^{\dagger}\right)/(2i)$, and $\dagger$ denotes the conjugate transpose operator. Here, the matrix $\underline{\boldsymbol{\varepsilon}}''$ is  Hermitian, ensuring that the computed dissipated power is real-valued. If the dissipated power is negative, the material transfers energy to the optical field, resulting in amplification. Throughout the remainder of this paper, we use normalized dynamic electric fields with an amplitude of 1 V/m when evaluating the dissipated power. \textcolor{black}{The gyrotropic part of the effective permittivity tensor does not contribute to power exchange, because $\underline{\boldsymbol{\varepsilon}}_{\rm EO}^{\rm G}{''} = \left(\underline{\boldsymbol{\varepsilon}}_{\rm EO}^{\rm G} - \underline{\boldsymbol{\varepsilon}}_{\rm EO}^{\rm G}{}^{\dagger}\right)/(2i)=0$; it only contributes to nonreciprocity.}

In the following sections, we investigate three specific cases by varying both the wave propagation direction and the orientation of the applied static electric field in biased $n$-doped Te, \textcolor{black}{and we aim at determining modes' polarization and the biasing threshold for a mode to amplify in each case.}

\subsection{Case I: Both static bias and wave propagation are along $z$}

In this case, we assume that both the static electric bias and wave propagation are aligned along the trigonal axis (the $z$-axis), i.e., $\mathbf{E_0} = E_{0} \hat{\mathbf{z}}$ and $\mathbf{k} = k \hat{\mathbf{z}}$, as shown in Fig. \ref{fig:generalViewofCases}(b). Based on Eq. (\ref{eq:GeneralDispersion}), the dispersion relation reduces to
\begin{equation}
    \det
    \begin{bmatrix}
        k^2 - k_0^2 \varepsilon_{\mathrm{\perp}} & - k_0^2 \varepsilon_{\mathrm{EO}, xy} & 0 \\
        - k_0^2 \varepsilon_{\mathrm{EO}, yx} & k^2 - k_0^2 \varepsilon_{\mathrm{\perp}} & 0\\
         0 & 0 & - k_0^2 \varepsilon_{\mathrm{||}}
    \end{bmatrix}
    = 0.
    \label{eq:CaseIDispersion}
\end{equation}
The resulting wavenumbers and their associated polarization states are
\begin{subequations}
\begin{equation}
 k_{1,2}=k_0 \sqrt{\varepsilon_{1,2}}, 
\end{equation}
\begin{equation}
    \mathbf{E}_{1,2} = 
	\begin{bmatrix}
	1\\i \\0
	\end{bmatrix}, 
	\begin{bmatrix}
    1\\-i\\0
    \end{bmatrix},
    \label{eq:PolCase1}
\end{equation}
\end{subequations}
\noindent where $\varepsilon_{1,2} = \varepsilon_{\mathrm{\perp}} \pm i \varepsilon_{\mathrm{EO}, xy}$ represent the effective permittivities for the two circularly polarized modes. The corresponding dissipated power density for this case is given by
\begin{align}
p_{\mathrm{dis}} 
&= \frac{1}{2} \varepsilon_0 \left( \omega \varepsilon^{''}_{\mathrm{d, \perp}} + \frac{ \Gamma \omega_{\mathrm{p, \perp}}^2}{\omega^2 + \Gamma^2} \right) \left( |E_x|^2 + |E_y|^2 \right) \notag\\
&+\frac{1}{2}  \varepsilon_0 \left( \frac{ \Gamma \omega_{\mathrm{c}, z}   \omega}{\omega^2 + \Gamma^2 } \right) \Im{(E_x E_y^*)}.
\label{eq:CaseIdissipatedPower}
\end{align}
The first term in Eq. (\ref{eq:CaseIdissipatedPower}), arising from the imaginary part of the diagonal elements of the permittivity tensor $\underline{\boldsymbol{\varepsilon}}$, is always positive, as Tellurium's dielectric response is inherently lossy in the frequency range from 0.1 THz to 30 THz. However, the second term, due to the non-Hermitian part associated with the off-diagonal elements of $\underline{\boldsymbol{\varepsilon}}$, can become negative. Only the second term on the right-hand side of Eq. (\ref{eq:EOGeneralConductivity}) is responsible for gain. The first part of Eq. (\ref{eq:EOGeneralConductivity}), though depending on the relaxation rate $\tau$, leads to a Hermitian component of the permittivity tensor $\underline{\boldsymbol{\varepsilon}}$ because of the gyrotropic effect and does not contribute to either losses or gain.

\begin{figure*}[bt!]
	\centering
	\includegraphics[scale=0.5]{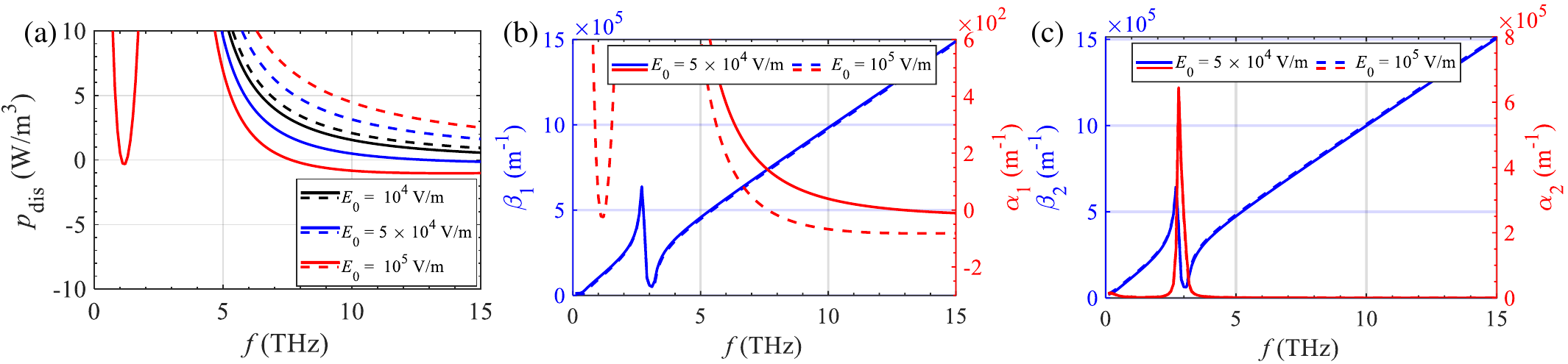} 
	\caption{(a) Dissipated power density for Case I, showing results for mode 1 (solid curves) and mode 2 (dashed curves). Negative values indicate optical amplification. For a bias of $E_0 = 5 \times 10^4 \;\mathrm{V/m}$, amplification of mode 1 occurs for $f > 12.95\;\mathrm{THz}$. Increasing the bias to $E_0 = 10^5 \;\mathrm{V/m}$ broadens the amplification range to $1.05\ \mathrm{THz} < f < 1.27\ \mathrm{THz}$ and for $f > 7.65\;\mathrm{THz}$. (b) Real and imaginary parts of the wavenumber $k_1$ for different values of the static electric bias. The zero crossings of $\alpha_1$ align exactly with the amplification regions shown in (a). (c) Real and imaginary parts of the wavenumber $k_2$ for different values of the static electric bias. Both $\beta_2$ and $\alpha_2$ remain positive across all frequencies, indicating that this mode is always decaying.} 
	\label{fig:CaseI_amplification}
\end{figure*}

Since the material is $n$-doped and the BCD parameter $D$ is negative, the resulting $\omega_{\mathrm{c}, z}$ becomes positive when $E_{0,z}>0$. The sign of $\omega_{\mathrm{c}, z}$ is reversed when the bias $E_{0,z}$ is reversed, and this is true also for the next two cases when their respective biases are reversed, as follows from Eq. \eqref{eq:cyclotron_frequency}. Beyond doping, previous studies, such as \cite{sinha2022berry}, have shown that the sign of BCD can be reversed through applied voltages, enabling identification of topological transitions. As $\omega \to 0$ or $\omega \to \infty$, the dissipated power $p_{\mathrm{dis}}$ remains positive. However, for mode 1, where $\Im \left(E_x E_y^* \right) <0$, the second term in Eq. (\ref{eq:CaseIdissipatedPower}) becomes negative, leading to net negative dissipated power over at least one frequency interval, indicating optical amplification as shown with solid curves in Fig. \ref{fig:CaseI_amplification}(a). As expected, increasing the bias field $E_0$ leads to stronger amplification and shifts the onset of amplification to lower frequencies. In this case, the minimum bias required to observe amplification is \textcolor{black}{$2.03 \times 10^4 \;\mathrm{V/m}$ at 30 THz.} Considering mode 2, where $\Im \left(E_x E_y^* \right) > 0$, the dissipated power is always positive, as illustrated by the dashed curves in Fig. \ref{fig:CaseI_amplification}(a), and the mode is attenuating. \textcolor{black}{In summary, it is mode 1 that is subject to amplification, and it can be left-handed or right-handed circularly polarized depending on the direction of propagation, a clear sign of anisotropy induced by the gyrotropy of the medium.}

To better understand the mode amplification mechanism, we calculated the real part of the Poynting vector, defined as $\mathbf{S}_r = \Re (\mathbf{E} \times \mathbf{H}^*)/2$ for the two modal solutions, given by $\mathbf{E}= \mathbf{E}_{1,2} e^{\pm i k_{1,2} z}$, where the $\pm$ sign for each mode accounts for propagation in the $\pm z$ directions. From $\nabla \times \mathbf{E} = i \omega \mu_0 \mathbf{H}$, we obtain the corresponding magnetic field as $\mathbf{H} = \pm \hat{\mathbf{z}} \times \mathbf{E}_n k_n/(\omega \mu_0) e^{\pm i k_n z}$, where $n=1,2$ labels the mode numbers. Substituting into the Poynting vector expression yields,
\begin{equation}
    \mathbf{S}_{r,n} = \pm \frac{1}{2} \frac{ \beta_n}{\omega \mu_0} |\mathbf{E}_n|^2 e^{\mp 2 \alpha_n z} \hat{\mathbf{z}}.
    \label{eq:PoyntingVector}
\end{equation}
\noindent For mode 1, $\alpha$ is negative while $\beta$ is positive, as illustrated in Fig. \ref{fig:CaseI_amplification}(b). This indicates the Poynting vector, obtained in Eq. \eqref{eq:PoyntingVector}, confirms that this mode grows in the same direction of the power flow, regardless of whether it propagates in the $+z$ or $-z$ direction. In contrast, for mode 2, both $\alpha$ and $\beta$ have the same sign, as shown in Fig. \ref{fig:CaseI_amplification}(c), which means that the mode decays in the direction of power flow, again independent of the propagation direction. In this mode, $\alpha$ and $\beta$ always have the same sign. In summary, mode 1 experiences amplification regardless of the direction of propagation, while mode 2 always undergoes attenuation. It is important to note that the amplifying mode 1 reverses handedness depending on the direction of propagation.

If the direction of the applied static bias is reversed, the sign of $E_{0,z}$ changes, which in turn flips the sign of $\omega_{\mathrm{c},z}$. As a result, mode 2 becomes the amplifying mode, while mode 1 experiences loss. Unlike conventional lasers, which typically require a stabilization mechanism to fix the polarization of emitted light, this system enables dynamic control over the polarization handedness by reversing the direction of the static bias. This behavior arises from the chiral nature of BCD-induced gain, allowing straightforward switching between left- and right-handed amplified modes.

\subsection{Case II: Static bias along $z$ and wave traveling  along $x$}
\begin{figure*}[bt!]
	\centering
	\includegraphics[scale=0.50]{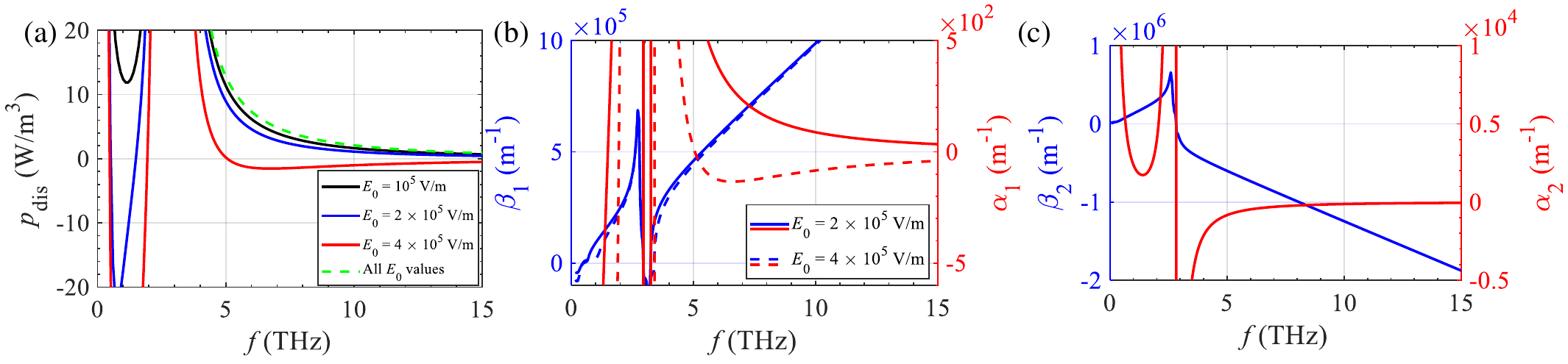} 
	\caption{(a) Dissipated power density for Case II, showing results for mode 1 (solid curves) and mode 2 (green dashed curves). Negative values indicate optical amplification. For a bias of $E_0 = 2\times10^5\;\mathrm{V/m}$, mode 1 exhibits amplification in the range $0.50\ \mathrm{THz} < f < 1.49\ \mathrm{THz}$. Increasing the bias to $E_0 = 4\times10^5\;\mathrm{V/m}$ expands the amplification range to $0.45\;\mathrm{THz} < f < 1.95\;\mathrm{THz}$ and for $f > 5.06\ \mathrm{THz}$. The dissipated power for mode 2 remains strictly positive and is independent of the applied static bias. (b) Real and imaginary parts of the wavenumber $k_1$ for different values of the static electric bias. (c) Real and imaginary parts of the wavenumber $k_2$, which do not depend on $E_0$.} 
	\label{fig:CaseII_amplification}
\end{figure*}

In this case, we assume that the static electric bias is aligned along the trigonal axis (the $z$-axis) and the wave propagates along the $x$-axis, orthogonal to the trigonal axis, i.e., $\mathbf{E_0} = E_{0} \hat{\mathbf{z}}$ and $\mathbf{k} = k \hat{\mathbf{x}}$, as shown in Fig. \ref{fig:generalViewofCases}(c). Based on Eq. (\ref{eq:GeneralDispersion}), the dispersion relation reduces to
\begin{equation}
    \det
    \begin{bmatrix}
         - k_0^2 \varepsilon_{\mathrm{\perp}} & - k_0^2 \varepsilon_{\mathrm{EO}, xy} & 0\\
        - k_0^2 \varepsilon_{\mathrm{EO}, yx} & k^2 - k_0^2 \varepsilon_{\mathrm{\perp}} & 0\\
         0 & 0 & k^2- k_0^2 \varepsilon_{\mathrm{||}}
    \end{bmatrix}
    = 0.
    \label{eq:CaseIIDispersion}
\end{equation}
The wavenumbers and their associated polarization states are 
\begin{subequations} 
	\begin{equation}
	k_{1}=k_0 \sqrt{\frac{\varepsilon_{\mathrm{\perp}}^2 + \varepsilon_{\mathrm{EO}, xy}^2}{\varepsilon_{\mathrm{\perp}}}}, \quad
    k_2 = k_0 \sqrt{\varepsilon_{\mathrm{||}}},
    \label{eq:CaseIIEigenvalues}
	\end{equation}  
        \begin{equation}
	\mathbf{E}_{1} = 
	\begin{bmatrix}
	-\dfrac{ \varepsilon_{\mathrm{EO}, xy}}{\varepsilon_{\mathrm{\perp}}}\\1 \\0
	\end{bmatrix}, \quad
    \mathbf{E}_{2} =
	\begin{bmatrix}
    0\\0\\1
    \end{bmatrix}.
    \label{eq:CaseIIEigenvectors}
	\end{equation}
\end{subequations}
The corresponding dissipated power density for mode 1 in this case is still given by Eq. \eqref{eq:CaseIdissipatedPower}. 
%
%
Figure \ref{fig:CaseII_amplification}(a) shows that \textcolor{black}{the first frequency at which} the dissipated power density becomes negative is \textcolor{black}{0.86 THz}, which corresponds to mode 1 at the threshold bias of $E_0 \geq 1.41 \times 10^5\;\mathrm{V/m}$. For a larger bias, $E_0 = 2 \times 10^5\;\mathrm{V/m}$, mode 1 exhibits amplification within the frequency range $0.50\ \mathrm{THz} < f < 1.45\ \mathrm{THz}$. Also, when the bias is increased to $E_0 = 4 \times 10^5\;\mathrm{V/m}$, amplification occurs for $0.45\ \mathrm{THz} < f < 1.95\ \mathrm{THz}$ and for $f > 5.06\;\mathrm{THz}$. The corresponding real and imaginary parts of the wavenumber for the eigenmodes in Case II are plotted in Fig. \ref{fig:CaseII_amplification}(b), confirming the frequency intervals associated with gain. Furthermore, the amplification of mode 1 is also verified by evaluating the Poynting vector, as given in Eq. \eqref{eq:PoyntingVector}, following the same approach used in Case I.  As for Case I, the amplifying mode 1 reverses handedness depending on the direction of propagation.  

Additionally, the corresponding dissipated power density for mode 2 is given by
\begin{equation}
p_{\mathrm{dis,2}} = \frac{1}{2} \varepsilon_0 \left( \omega \varepsilon^{''}_{\mathrm{d, ||}} + \frac{ \Gamma \omega_{\mathrm{p, ||}}^2}{\omega^2 + \Gamma^2} \right) |E_z|^2, \\
\label{eq: CaseIIDissipatedPowerMode2}
\end{equation}
which is always positive, indicating that mode 2 is attenuating across the entire frequency range. For mode 2, both the dissipated power and the wavenumber depend solely on frequency and are independent of the applied static bias, as illustrated by the green dashed curves in Fig. \ref{fig:CaseII_amplification}(a) and the plots in Fig. \ref{fig:CaseII_amplification}(c), respectively.

\begin{figure}[t!]
	\centering
	\includegraphics[scale=0.47]{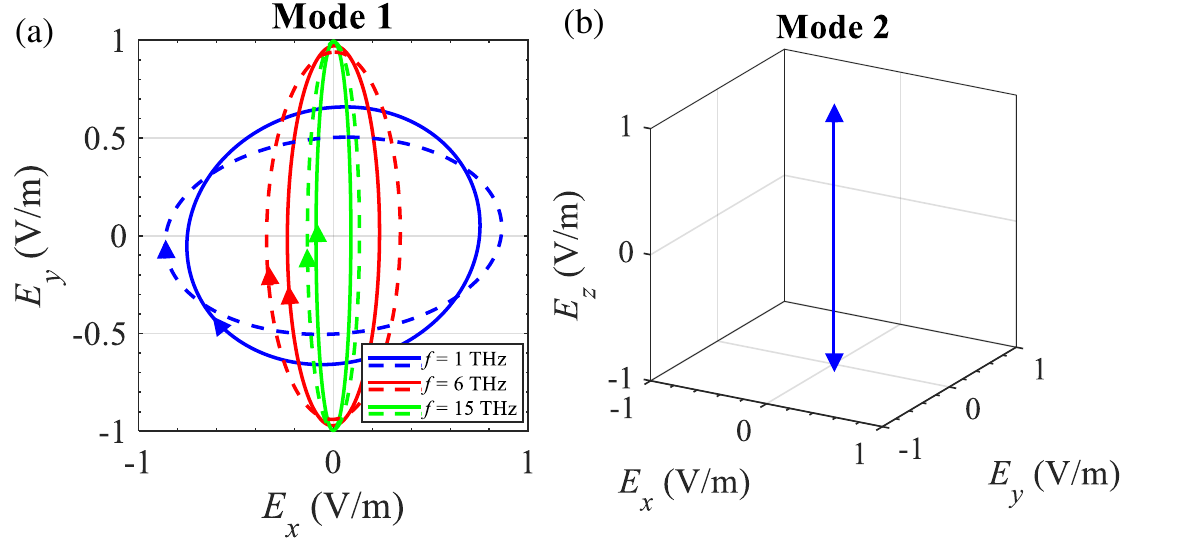} 
	\caption{Polarization eigenstates for Case II: (a) Amplifying mode (mode 1) at three representative frequencies, showing elliptical polarization. Results are plotted for two static electric field biases, $E_0 = 4 \times 10^5\;\mathrm{V/m}$ (solid curves) and $E_0 = 6 \times 10^5\;\mathrm{V/m}$ (dashed curves). (b) Decaying mode (mode 2) exhibits linear polarization aligned along the $z$-axis.} 
	\label{fig:CaseII_polarization}
\end{figure}

In this case, mode 1 is amplifying, and the electric field includes the longitudinal component $E_x$ despite the wavevector being purely along $x$. It exhibits elliptical polarization with electric spin orthogonal to the direction of propagation; its ellipticity can be tuned by adjusting the magnitude of the applied static electric field bias $E_0$, as shown in Fig. \ref{fig:CaseII_polarization}(a). Furthermore, the handedness can be reversed by inverting the bias $E_0$ because this affects the sign of the effective cyclotron-like angular frequency $\omega_{\mathrm{c}, z}$. 

In contrast, mode 2 is always attenuating and maintains a linear polarization strictly along the $z$-axis, independent of the bias strength, as shown in Fig. \ref{fig:CaseII_polarization}(b).

\subsection{Case III: Both static bias and wave propagation are along $x$}

In this case, we assume that both the static electric bias and wave propagation are aligned along the $x$-axis, orthogonal to the trigonal axis, i.e., $\mathbf{k} = k \mathbf{\hat{x}}$ and $\mathbf{E_0} = E_0 \hat{\mathbf{x}}$ as shown in Fig. \ref{fig:generalViewofCases}(d). Based on Eq. \eqref{eq:GeneralDispersion}, the dispersion relation reduces to
\begin{equation}
    \det
    \begin{bmatrix}
        - k_0^2 \varepsilon_{\mathrm{\perp}} & 0 & 0 \\
        0 & k^2 - k_0^2 \varepsilon_{\mathrm{\perp}} & - k_0^2 \varepsilon_{\mathrm{EO}, yz}\\
         0 & - k_0^2 \varepsilon_{\mathrm{EO}, zy} & k^2 - k_0^2 \varepsilon_{\mathrm{||}}
    \end{bmatrix}
    = 0.
    \label{eq: Xdispersion}
\end{equation}
The wavenumbers and their associated polarization states are 
\begin{subequations}
\begin{equation}
 k_{1,2}=k_0 \sqrt{\varepsilon_{1,2}},
\end{equation}    
\begin{equation}
	\mathbf{E}_{1, 2} = 
\begin{bmatrix}
	0\\
	\dfrac{  \varepsilon_{1,2}-\varepsilon_{\mathrm{||}}}{ \varepsilon_{\mathrm{EO}, zy}}\\
	1
\end{bmatrix},
\label{eq:caseIIIPolar}
\end{equation}
\end{subequations}
where the effective permittivities for the two eigenmodes are 
\begin{equation}
	\varepsilon_{1, 2} = \frac{1}{2} \left( \varepsilon_{\mathrm{\perp}} + \varepsilon_{\mathrm{||}} \pm \sqrt{(\varepsilon_{\mathrm{\perp}} - \varepsilon_{\mathrm{||}})^2 + 4 \varepsilon_{\mathrm{EO}, yz} \varepsilon_{\mathrm{EO}, zy} }\right).
    \label{eq:CaseIII_eps1,2}
\end{equation}

Wave propagation along Te’s helical chains (Case I), under a static electric field bias applied in the same direction, experiences a permittivity tensor whose transverse components are equal in magnitude and phase-shifted by $\pi/2$, resulting in perfect circular polarization. In contrast, wave propagation orthogonal to the chains, as in Case II and this case, experiences unequal transverse permittivities, which distort the circular polarization into an elliptical polarization. The degree of ellipticity depends on the applied DC bias.

\begin{figure*}[t!]
	\centering
	\includegraphics[scale=0.5]{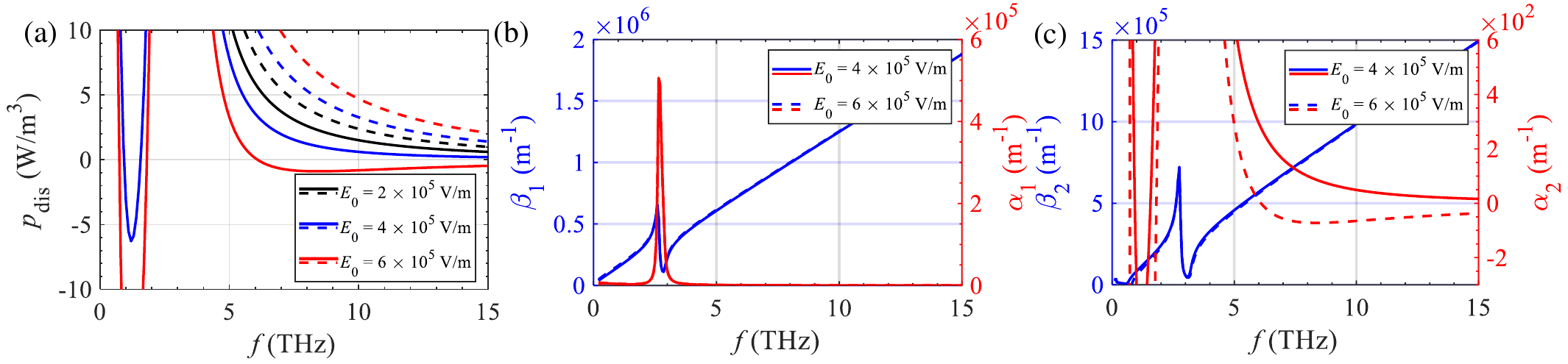} 
	\caption{(a) Dissipated power density for Case III, showing results for mode 2 (solid curves) and mode 1 (dashed curves). Negative values indicate optical amplification. For a bias of $E_0 = 4 \times 10^5 \;\mathrm{V/m}$, mode 2 exhibits amplification in the range $0.9\;\mathrm{THz} < f < 1.6\;\mathrm{THz}$. Increasing the bias to $E_0 = 6 \times 10^5 \;\mathrm{V/m}$ expands the amplification range to $0.7\;\mathrm{THz} < f < 1.8\ \mathrm{THz}$ and for $f > 6\;\mathrm{THz}$. (b) Real and imaginary parts of the wavenumber $k_1$ for different values of the static electric bias. Both $\beta_1$ and $\alpha_1$ remain positive across all frequencies, indicating that this mode is always decaying. (c) Real and imaginary parts of the wavenumber $k_2$ for different values of the static electric bias.} 	\label{fig:CaseIII_amplification}
\end{figure*}

The energy transfer from the material to each polarization eigenstate is evaluated using the dissipated power density,
\begin{align}
p_{\mathrm{dis}} 
&= \frac{1}{2} \varepsilon_0 \left( \omega \varepsilon^{''}_{\mathrm{d, \perp}} + \frac{ \Gamma \omega_{\mathrm{p, \perp}}^2}{\omega^2 + \Gamma^2} \right) |E_y|^2 \notag \\
&+\frac{1}{2} \varepsilon_0 \left( \omega \varepsilon^{''}_{\mathrm{d, ||}} + \frac{ \Gamma \omega_{\mathrm{p, ||}}^2}{\omega^2 + \Gamma^2} \right) |E_z|^2 \notag \\
&+ \frac{3}{2} \varepsilon_0  \frac{\Gamma^2\omega_{\mathrm{c}, x}}{\omega^2 + \Gamma^2} \Re({E_y^* E_z}) \notag\\
&- \frac{1}{2} \varepsilon_0   \frac{\Gamma \omega_{\mathrm{c}, x} \omega}{\omega^2 + \Gamma^2} \Im({E_y^* E_z}), 
\label{eq: xDissipatedPower}
\end{align}
which can become negative for mode 2 under a sufficiently large static electric field bias, indicating optical amplification. In this scenario, the threshold bias necessary to observe amplification is roughly $ 3.26 \times 10^5 \;\mathrm{V/m}$ \textcolor{black}{at 1.23 THz}. Once the applied electric field exceeds this value, the dissipated power becomes negative, as shown in Fig. \ref{fig:CaseIII_amplification}(a).

As shown in Fig. \ref{fig:CaseIII_amplification}(a), mode 1 consistently exhibits decay across the entire frequency range. Consequently, both the real and imaginary parts of its wavenumber share the same sign, as illustrated in Fig. \ref{fig:CaseIII_amplification}(b). In contrast, for electric field biases exceeding the amplification threshold, mode 2 displays frequency intervals in which the real and imaginary parts of the wavenumber have opposite signs, indicating optical amplification. These regions are clearly visible in Fig. \ref{fig:CaseIII_amplification}(c).

\begin{figure}[t!]
	\centering
	\includegraphics[scale=0.48]{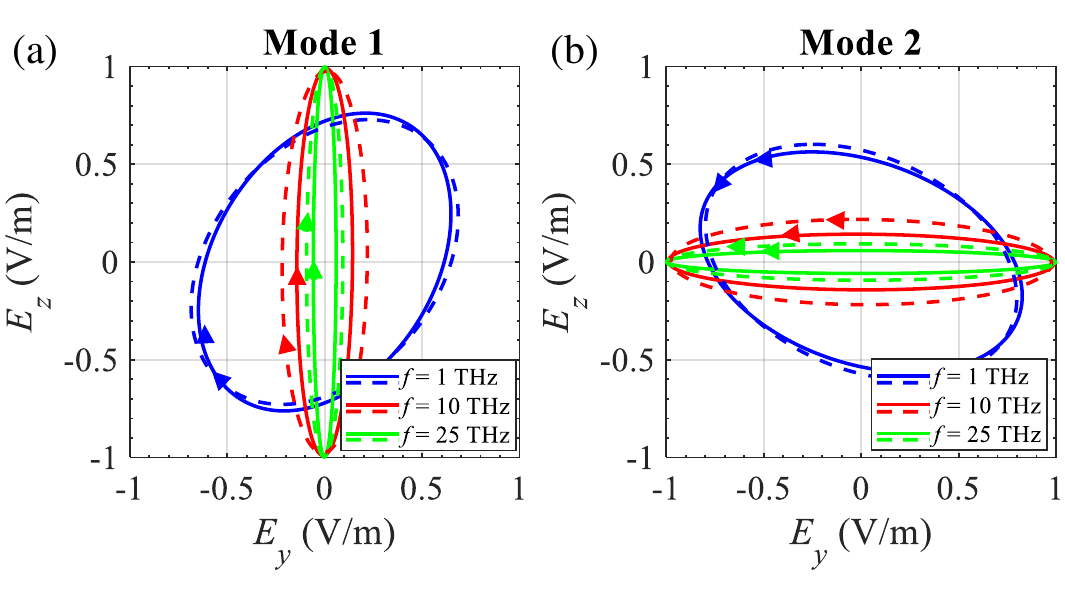} 
	\caption{Polarization eigenstates for Case III: (a) Decaying mode (mode 1) and (b) Amplifying mode (mode 2) at three representative frequencies, for two static electric field biases $E_0 = 5 \times 10^5 \;\mathrm{V/m}$ (solid curves) and $E_0 = 8 \times 10^5 \;\mathrm{V/m}$ (dashed curves). } 
	\label{fig: ExPolarization}
\end{figure}

In this scenario, similar to Case I, we observe chiral eigenstates; however, the polarizations are elliptical rather than circular. In contrast to the results of \cite{hakimi2024chiral}, where elliptical eigenstates appeared at all frequencies regardless of the applied bias, here the ellipticity depends on both the operating frequency and the static electric field. Figure \ref{fig: ExPolarization} illustrates the two polarization eigenstates at three representative frequencies for bias values of $E_0 = 5 \times 10^5\:\mathrm{V/m}$ (solid curves) and $E_0 = 8 \times 10^5\:\mathrm{V/m}$ (dashed curves). As frequency increases, the elliptical nature of the eigenstates becomes more pronounced, and at a fixed frequency, the degree of ellipticity can be tuned via the applied electric field.

\textcolor{black}{A dynamic bias-control of elliptical polarization is desirable for real-time polarization modulation in terahertz sensing and communication systems \cite{huang2021polarization}. For instance, in \cite{zaman2023versatile}, the ellipticity of the transmitted THz wave is tuned by electrostatic gating of the graphene layer. This gate bias changes the graphene’s conductivity, giving a continuous ellipticity modulation in a limited range. Furthermore, in \cite{kindness2020terahertz}, active control of circular dichroism and optical activity around 2 THz was achieved with a double-layer ring-resonator metamaterial incorporating an electrostatically gated graphene monolayer. Such functionality is suitable for polarization-sensitive spectroscopy, imaging, and active manipulation of QCL outputs for compact amplitude/polarization modulators. Similarly, reference \cite{song2024thz} fabricated a patterned graphene–gold bilayer metasurface and experimentally showed that the applied bias voltage electrically tunes the reflected polarization, including both ellipticity and rotation angle. Their experimental measurements highlight the potential for THz communications and sensing.}

\begin{figure}[t!]
	\centering
	\includegraphics[scale=0.38]{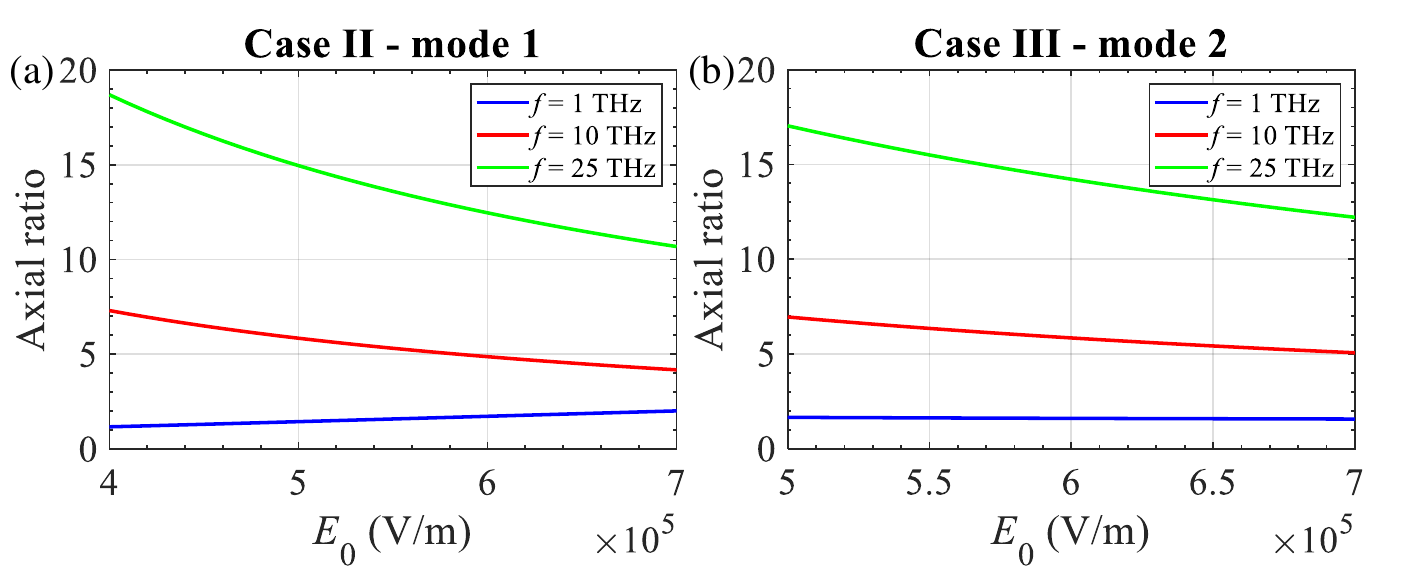} 
	\caption{\textcolor{black}{Axial ratio for the amplifying modes of (a) Case II, mode 1, and (b) Case III, mode 2. In both plots, the range of $E_0$ is chosen high enough to ensure amplification at the selected frequencies.}} 
	\label{fig:AR_amplification}
\end{figure}

\begin{figure}[t!]
	\centering
	\includegraphics[scale=0.38]{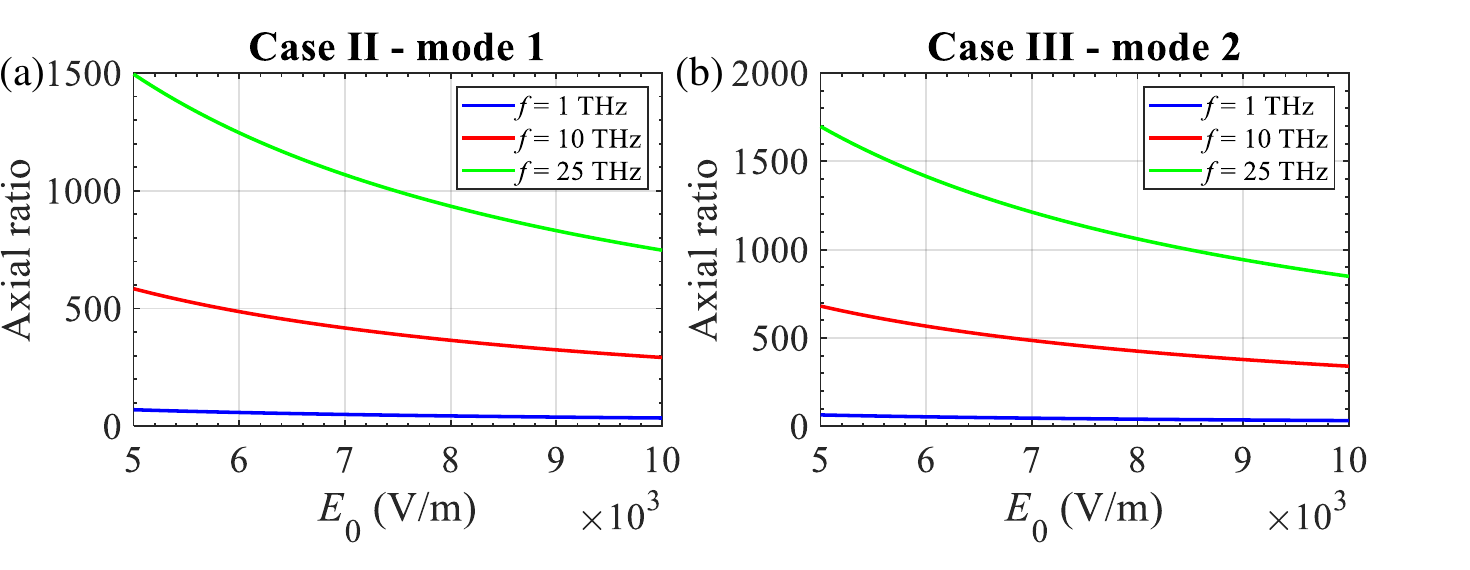} 
	\caption{\textcolor{black}{As in Fig.~\ref{fig:AR_amplification} but for smaller $E_0$ values such that the modes experience no amplification, but the axial ratio is still bias dependent.}} 
	\label{fig:AR_loss}
\end{figure}

\textcolor{black}{Consistent with the goal of previously mentioned studies, in Cases II and III of this work, we show how electrical bias modifies ellipticity. Ellipticity tuning for different values of $E_0$ is evaluated using the axial ratio definition in \cite{balanis2016antenna}. Fig. \ref{fig:AR_amplification} shows the axial ratio across different frequencies, assuming the bias values produce amplification in Cases II and III. Notably, the bias-dependent ellipticity does not require modal gain; it also occurs in lossy operating regimes, as illustrated in Fig. \ref{fig:AR_loss}. Since the ellipticity modulation occurs even in lossy regimes, these effects may be experimentally accessible through polarization-resolved transmission or reflection measurements without requiring gain or lasing.}

 \subsection{Summary of the three cases}
 
\begin{figure}[t!]
	\centering
	\includegraphics[scale=0.18]{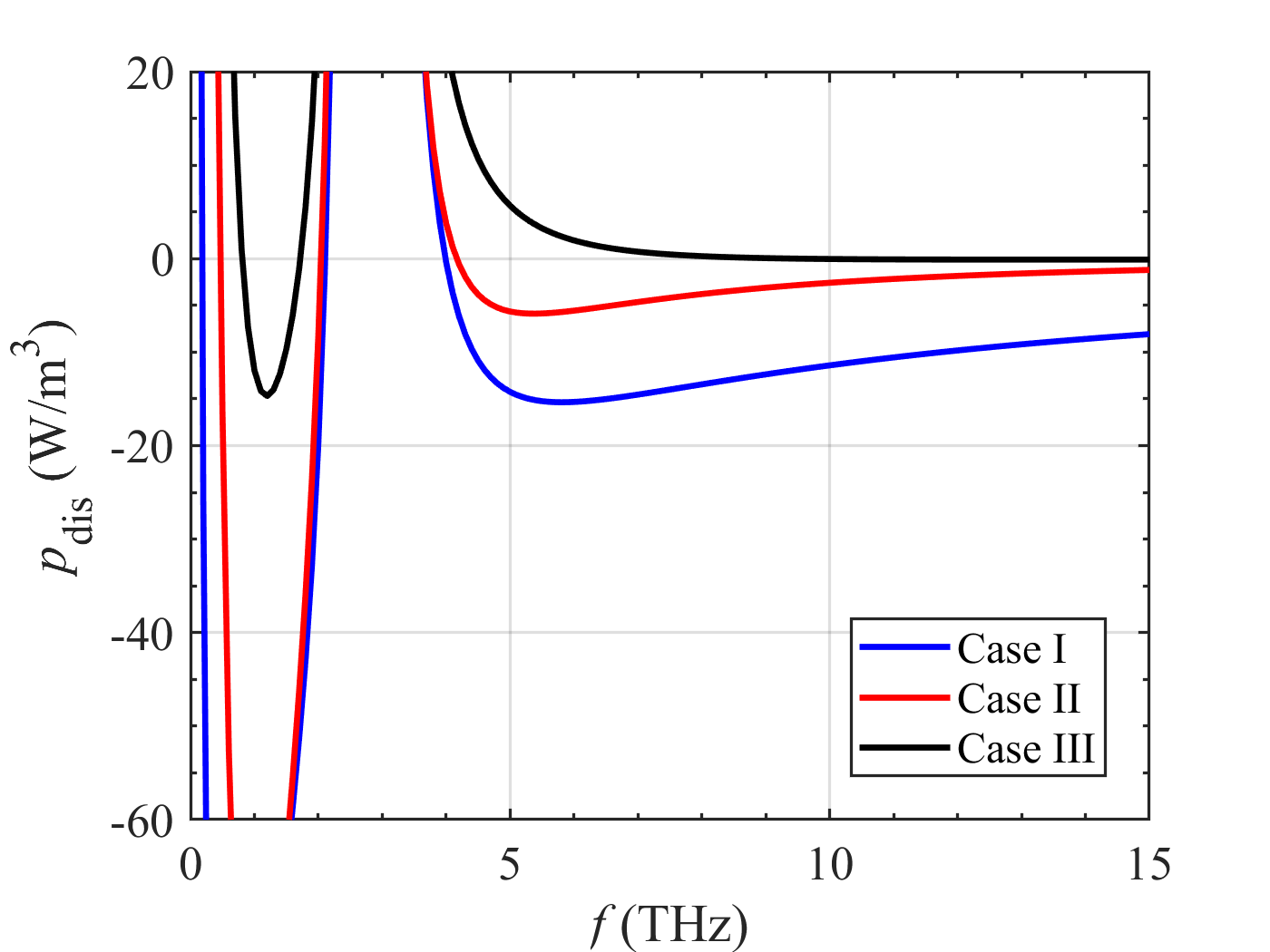} 
	\caption{Dissipated power for the growing mode in each of the three cases studied, assuming a bias DC voltage of $E_0 = 5 \times 10^5 \;\mathrm{V/m}$. Case I shows the strongest amplification per unit length.} 
	\label{fig:DissipatedPowerComparison}
\end{figure}

 To highlight the difference in amplification behavior in these three cases, Fig. \ref{fig:DissipatedPowerComparison} compares the dissipated power for all three cases at a fixed bias of $E_0 = 5\times 10^5 \;\mathrm{V/m}$. 

To better contextualize our findings, it is instructive to qualitatively compare them with those reported in \cite{hakimi2024chiral}, where twisted bilayer graphene served as the active medium for amplification. Although achieving comparable levels of amplification in our bulk 3D Tellurium system requires a higher electric bias, it is important to emphasize that Tellurium’s bulk nature avoids the complex fabrication challenges associated with stacking multiple twisted bilayers of graphene \cite{nimbalkar2020opportunities}. This inherent simplicity in material processing makes bulk Tellurium a promising and more accessible platform for terahertz amplification and lasing.

Recent studies have demonstrated that cavities made from low-symmetry two-dimensional structures can exhibit chiral lasing driven by the BCD gain \cite{hakimi2024chiral}. In the next section, we show that Tellurium, in the three cases discussed earlier, can serve as an active medium in a micrometer-scale cavity to design a simple chiral laser.

\section{Chiral THz Lasing using Bulk \(\bm{n}\)-doped Tellurium}
\label{Sec:lasing}
A terahertz laser is conceived by placing biased Tellurium within a cavity formed by two mirrors separated by a distance $L$. \textcolor{black}{In all three cases studied,  when the wave inside the cavity reflects from a mirror, the handedness of its polarization is reversed. Due to the change of propagation direction after reflection, there is a sign flip in the wavevector in  Eq. \eqref{eq:WaveEquation_k},  $\mathbf{k}\to-\mathbf{k}$, but it does not affect the polarization states in Eqs. \eqref{eq:PolCase1}, \eqref{eq:CaseIIEigenvectors}, and \eqref{eq:caseIIIPolar}, hence it does not affect the equations of the amplification analysis. As a result, despite the reversal of the handedness upon reflection, the gyrotropy-induced nonreciprocity causes that it is the same polarization state  (i.e., an eigenvector in Eqs. \eqref{eq:PolCase1}, \eqref{eq:CaseIIEigenvectors}, and  \eqref{eq:caseIIIPolar}) that continues to be associated with amplification.
Thus, both the nonreciprocal effect induced by the material's gyrotropy and the gain associated with the non-Hermitian part of the EO effect jointly enable self-sustained oscillations in a cavity using the same amplifying mode.  This phenomenon is the key to chiral lasing, as was also shown in \cite{hakimi2024chiral}. For the lasing threshold analysis in this section, we only consider the amplifying mode, which remains amplified after reflection despite the handedness reversal. The mirrors are assumed to be non-depolarizing. In the following analysis, we focus on the threshold condition for a single amplifying cavity mode; multimode operation and gain competition are not considered.} 

We assume one mirror to be perfectly reflective, with a reflection coefficient of -1, while the other mirror is partially reflective, allowing power to be emitted from the cavity. Under these assumptions, the lasing condition is given by
\begin{equation}
R e^{-2\alpha L} e^{i(2\beta L - \phi)} = -1,
\label{eq:generalLasingCondition}
\end{equation}
where $R$ and $\phi$ are the magnitude and phase of the field reflection coefficient of the partial mirror. Here, we consider both $\alpha$ and $\beta$ of the amplifying mode to be functions of applied static bias $E_0$ and $\omega$. For simplicity, we assume that $\phi = \pi$. \textcolor{black}{We determine the pair $(\omega_\mathrm{r},E_0)$ that satisfies Eq. \eqref{eq:generalLasingCondition}, where $\omega_\mathrm{r}$ is oscillation frequencies that satisfies the resonance condition 
\begin{equation}
    \beta(\omega_\mathrm{r}, E_0) = m \pi/L,
    \label{eq:lasing_2}
\end{equation}
with $m=1,2,3,...$, denoting the longitudinal mode number of the lasing cavity. In this section, we first consider the $m=1$ fundamental mode of oscillation, whereas the case with $m>1$ is summarized at the end and detailed in Appendix \ref{App: Higher-order modes}.} To satisfy the magnitude requirement of Eq. \eqref{eq:generalLasingCondition}, the following condition must also hold:
\begin{equation}
    \alpha(\omega_\mathrm{r}, E_0) = \ln(R)/(2L).
    \label{eq:lasing_1}
\end{equation}
 In this section, we assume $R = 0.99$. It is clear that with higher reflectivity, a smaller bias is needed to initiate oscillations. Several structures can provide such a high reflectivity at THz frequencies \cite{badloe2017metasurfaces, yan2023switchable}. In the following subsections, we will analyze the lasing threshold for the three cases.

\subsection{Lasing in Case I}
\begin{figure*}[t!]
  \centering
  \includegraphics[scale=0.48]{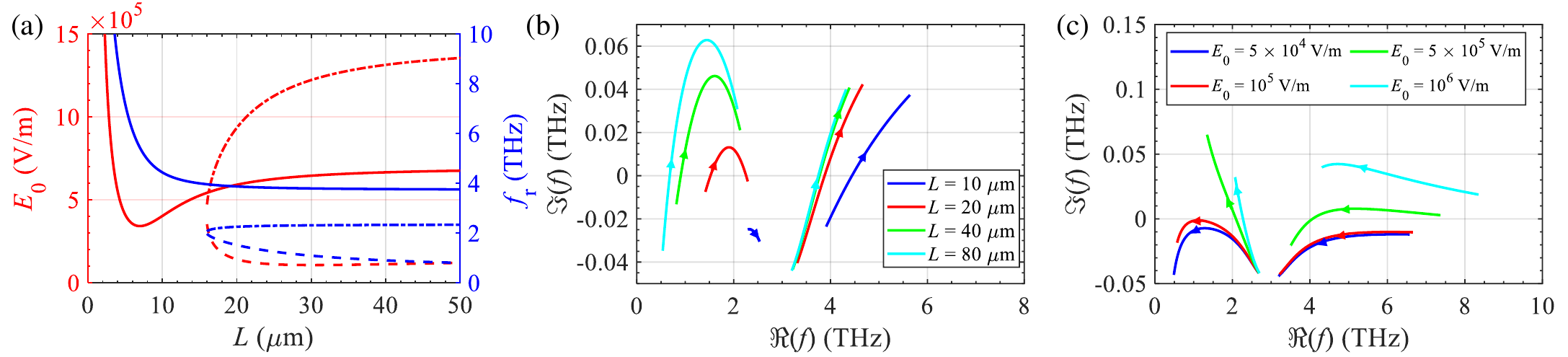} 
  \caption{Case I. (a) First fundamental oscillation frequency \textcolor{black}{($m=1$)} and corresponding threshold electric bias $E_0$ to achieve oscillations, varying cavity length $L$. The solid, dashed, and dash-dotted curves denote the three solution branches of Eqs. \eqref{eq:lasing_2} and \eqref{eq:lasing_1}. The minimum required bias is $1.07 \times 10^5 \;\mathrm{V/m}$, occurring at $L = 30.2 \;\mu \mathrm{m}$. (b) Imaginary part versus real part of the complex lasing frequency for $ 5 \times 10^4 \;\mathrm{V/m} < E_0 < 10^6 \;\mathrm{V/m}$ at each branch. (c) Imaginary versus real part of the complex lasing frequency for $5\:\mathrm{\mu m} < L < 100\:\mathrm{\mu m}$ at each branch.} 
  \label{fig:CaseI_Lasing}
\end{figure*}

For cavity lengths $L \in [1\, \mathrm{\mu m},\:50\:\mathrm{\mu m}]$ in the $z$ direction, we solve Eqs. (\ref{eq:lasing_1}) and (\ref{eq:lasing_2}) to determine the required static electric field bias $E_0$ and the corresponding resonance frequency at each cavity length, as shown in Fig. \ref{fig:CaseI_Lasing}(a).
The solid, dashed, and dash-dotted curves represent the three solution branches of these two nonlinear equations. 
Notably, for $L > 15.9 \;\mu \mathrm{m}$, three distinct solutions appear, reflecting the effect of resonance at around 3 THz in the dielectric permittivity of  Te (as shown in \eqref{eq:DielectricPermittivityLorentz1} and \eqref{eq:DielectricPermittivityLorentz2}). The minimum required static field bias of $E_0 = 1.07 \times 10^5 \;\mathrm{V/m}$ occurs at $L = 30.2 \;\mathrm{\mu m}$, with a resonance frequency of $1.09\;\mathrm{THz}$. 
In the cavity length region where only one solution exists, the minimum bias increases to $3.42 \times 10^5\;\mathrm{V/m}$, occurring at $L = 7\;\mu \mathrm{m}$ with the oscillation frequency of $5.42\;\mathrm{THz}$. 

The results in Fig. \ref{fig:CaseI_Lasing}(a) provide the onset of lasing frequencies; however, actual amplification requires a positive imaginary part of the complex frequency, i.e., $\Im(\omega) > 0$. This condition is obtained through complex frequency analysis, with Figs. \ref{fig:CaseI_Lasing}(b) and (c) illustrating the lasing regions as functions of static electric field bias and cavity length, respectively. The zero crossings of $\Im(\omega)$ for each length in Fig. \ref{fig:CaseI_Lasing}(b) correspond to the fundamental oscillation frequencies obtained in Fig. \ref{fig:CaseI_Lasing}(a). For instance, in Fig. \ref{fig:CaseI_Lasing}(b), for $L = 20\;\mu \mathrm{m}$, there are three such crossings within the bias range $ 5 \times 10^4 \;\mathrm{V/m} < E_0 < 10^6 \;\mathrm{V/m}$, consistent with results seen in Fig. \ref{fig:CaseI_Lasing}(a). For larger cavity lengths, the third zero crossing becomes inaccessible within the plotted bias range, as it requires electric fields exceeding $10^6\;\mathrm{V/m}$. Furthermore, in Fig. \ref{fig:CaseI_Lasing}(c), as $L$ is varied from $5\;\mathrm{\mu m}$ to $100\;\mathrm{\mu m}$, two solution branches are observed for each value of the applied bias.

\subsection{Lasing in Case II}
A similar analysis for lasing in Case I can be applied to the hybrid amplifying mode in Case II.  When this mode hits a reflective boundary, the tangential field component $E_y$  must reverse sign due to reflection, and the wave vector also inverts. Consequently, according to Eq. (\ref{eq:WaveEquation_k}), $E_x$ must also change sign. Thus, there is a $\pi$ phase shift in the field components of the hybrid mode as assumed in the analysis of Eq. (\ref{eq:lasing_2}). As explained earlier, in Case II, reversing the direction of ${E}_0$ flips the sign of the $x$-component of the polarization state given in Eq. \eqref{eq:CaseIIEigenvectors}, thereby reversing the amplifying mode's handedness. 

\begin{figure*}[bt!]
  \centering
  \includegraphics[scale=0.48]{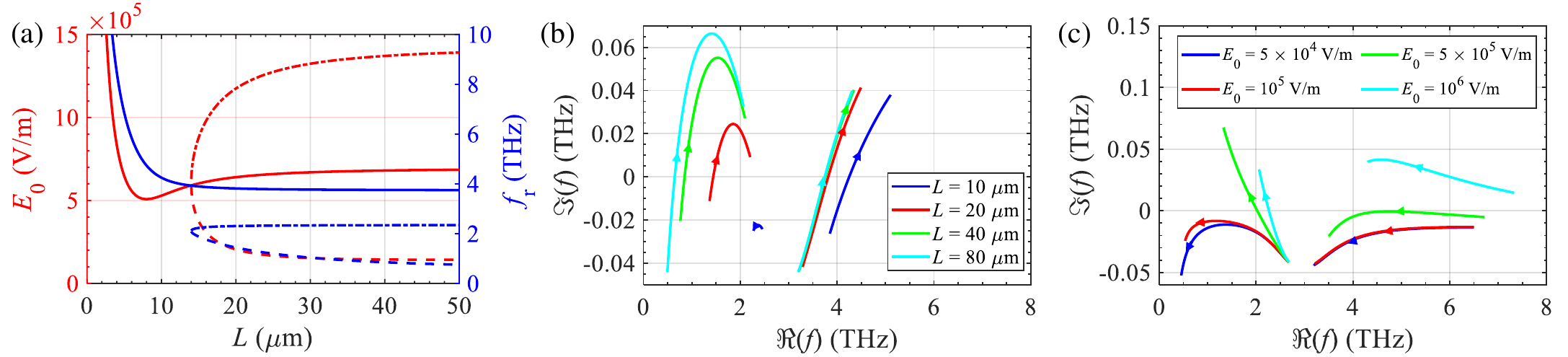} 
  \caption{Same as Fig. \ref{fig:CaseI_Lasing}, but for Case II, observing the following changes: (a) The minimum required bias is $1.42 \times 10^5 \;\mathrm{V/m}$, occurring at $L = 48.8 \;\mu m$.} 
  \label{fig:CaseII_Lasing}
\end{figure*}

Hence, for the lasing length in the $x$ direction, Fig. \ref{fig:CaseII_Lasing}(a) depicts the solutions of Eqs. (\ref{eq:lasing_1}) and (\ref{eq:lasing_2}) at each cavity length.
For $L < 14\;\mathrm{\mu m}$, the range in which we have a single solution, the minimum static electric field bias required is $5.08 \times 10^5\;\mathrm{V/m}$ at $L = 8\;\mathrm{\mu m}$. In the interval with three solutions, the minimum static electric field bias required is $E_0 = 1.42 \times 10^5 \;\mathrm{V/m}$ at $L = 48.8\;\mathrm{\mu m}$ with a resonance of $0.76\;\mathrm{THz}$. 

Using the same reasoning as in Figs. \ref{fig:CaseI_Lasing}(b) and (c), we obtain Figs. \ref{fig:CaseII_Lasing}(b) and (c). As it is clear, despite a slightly greater loss, this case produces imaginary values of $\omega$ very similar to those in Case I, indicating that its efficiency is roughly the same.
\subsection{Lasing in Case III}
\begin{figure*}[tb!]
  \centering
  \includegraphics[scale=0.48]{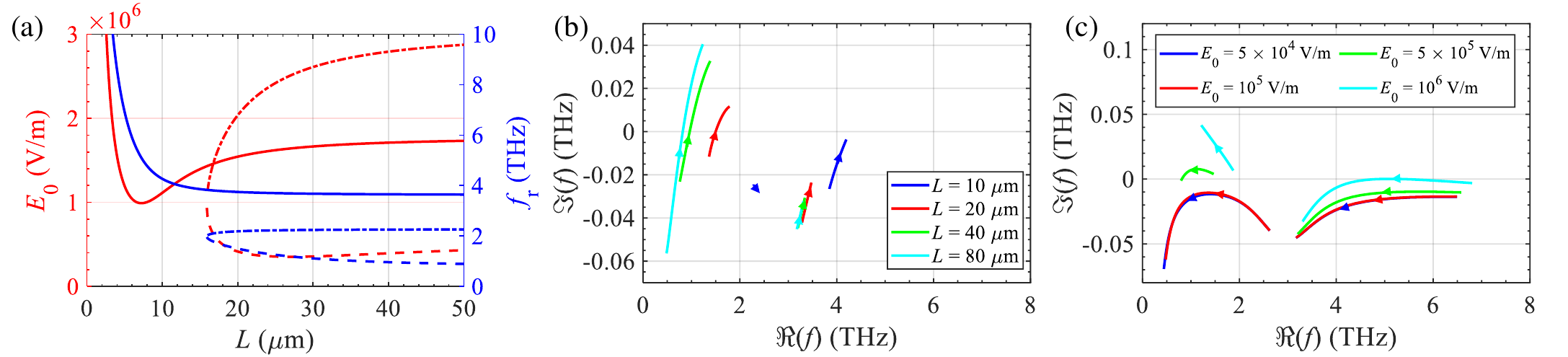} 
  \caption{Same as Fig. \ref{fig:CaseI_Lasing}, but for Case III, observing the following changes: (a) The minimum required bias is $3.52 \times 10^5 \;\mathrm{V/m}$, occurring at $L = 27.1 \;\mu m$.} 
  \label{fig:CaseIII_Lasing}
\end{figure*}

The solutions to Eqs. (\ref{eq:lasing_1}) and (\ref{eq:lasing_2}) as a function of cavity length $L$ in the $x$ direction are shown in Fig. \ref{fig:CaseIII_Lasing}(a). The minimum static electric field bias required for lasing occurs at $L = 27.1\: \mathrm{\mu m}$, with a threshold of $E_0 = 3.52 \times 10^5\;\mathrm{V/m}$. For smaller cavity lengths (corresponding to higher oscillation frequencies), the minimum required bias increases significantly. For example, at $L = 7.2\;\mathrm{\mu m}$, lasing is achieved at $E_0=9.95 \times 10^5\;\mathrm{V/m}$, with a oscillation frequency of $5.11\;\mathrm{THz}$. This suggests that operating at longer cavity lengths is preferable in Case III, as the required bias for shorter lengths approaches the material's breakdown field. Furthermore, a complex frequency analysis is performed to determine the instability behavior leading to lasing, with results shown in Fig. \ref{fig:CaseIII_Lasing}(b) when static electric field bias is varied and in Fig. \ref{fig:CaseIII_Lasing}(c) when cavity length is varied.

In this case, similar to Case II, inverting the sign of $E_0$ does not affect the $\varepsilon_{1,2}$ term in Eq. \eqref{eq:CaseIII_eps1,2}. Instead, only the EO response in the $zy$ direction, $\varepsilon_{\mathrm{EO}, zy}$, changes sign, which leads to the reversal of the handedness of the amplifying mode.

\subsection{Summary and Discussion}

To contextualize the required bias values, we briefly review experimentally measured breakdown and applied electric fields in Tellurium. Reference \cite{wang2006relationship} and  \cite{rudolph1971negative} report the breakdown electric field of intrinsic and $p$-type Tellurium are approximately $7.36 \times 10^5 \;\mathrm{V/m}$ and $7 \times 10^5 \;\mathrm{V/m}$ \cite{rudolph1971negative},  respectively. So, we expect the breakdown electric field of $n$-doped Te to be in the same range.  Furthermore, according to the estimation in \cite{morgado2024non}, the maximum $E_0$ applied in the experiment of \cite{shalygin2012current} was about $7 \times 10^3 \;\mathrm{V/m}$. Under special conditions, such as cooling and using a single crystal, a field bias up to $1.5 \times 10^5 \;\mathrm{V/m}$ has been applied to Tellurium without breakdown \cite{hoerstel1973high}. 

In our analysis, \textcolor{black}{considering only the first fundamental mode of oscillation ($m=1$)}, the minimum bias required to observe gain in Case I, assuming a laser cavity with mirrors of 99\% field reflectivity $(R=0.99)$, is approximately $1.07 \times 10^5 \;\mathrm{V/m}$, which is smaller than the breakdown value. For Cases II and III, the minimum required biases to observe gain are $1.42 \times 10^5 \;\mathrm{V/m}$ and $3.52 \times 10^5 \;\mathrm{V/m}$, respectively. Although these values are higher than those of Case I, they remain well below 50\% of the reported breakdown field, indicating that lasing operation is feasible within safe operating limits.

\textcolor{black}{Appendix \ref{App: Higher-order modes} presents the analysis of higher-order longitudinal modes of oscillations ($m>1$), and quantifies how increasing the mode number $m$ affects the oscillation frequency and the electric field bias required for lasing, at each cavity length. In addition, for every cavity length, we found the minimum bias required to initiate lasing. Considering all $m$ modes, in general, longer cavities further decrease the required biasing for lasing. However, in Case I, the oscillation frequency associated with the minimum required bias remains nearly unchanged from $L=30\ \mu\mathrm{m}$ to $L=57\,\mu\mathrm{m}$ and for $L>57\,\mu\mathrm{m}$ the oscillation frequency increases with length. In Cases II and III, over the range of $20\,\mu\mathrm{m}\leq L\leq100\,\mu\mathrm{m}$, the frequency that first reaches threshold stays quite constant with small fluctuations, rather than decreasing monotonically with increasing cavity length as expected for a conventional Fabry–Perot cavity. This is due to the emergence of branches of solutions when considering the Lorentz resonance at around 3 THz.}

\textcolor{black}{As a numerical illustration for a 100 $\mu$m cavity, Case I starts lasing for a bias value of $7.74\times10^{4}\ \mathrm{V/m}$ at $f_{\rm r}=16.78\ \mathrm{THz}$.}

\section{Conclusions}
\label{Sec:conclusions}
We have theoretically demonstrated that $n$-doped Te, a chiral non-centrosymmetric semiconductor,  supports optical amplification and lasing at THz frequencies enabled by its large BCD. By analyzing the non-Hermitian EO response under various configurations of static electric field orientation and optical field propagation direction, we identified three distinct regimes of circular, hybrid, and elliptical polarization modes, each capable of supporting an amplifying mode under appropriate conditions. Among the studied cases, the one where both static electric field and optical propagation are aligned along the trigonal $c$-axis (Case I) requires the lowest lasing threshold bias and yields circularly polarized amplification, making it particularly promising for practical THz device implementations. The other configurations enable dynamic control over polarization ellipticity, offering tunable chiral responses relevant to active devices. Furthermore, we demonstrate that micrometer-scale laser cavities incorporating Te as an active medium can reach lasing thresholds with electric bias well below the material’s breakdown limit. \textcolor{black}{Notably, for higher-order longitudinal modes ($m>1$), the electric bias required for lasing is further decreased. The lasing frequency at minimum threshold is not determined only by geometry but also by the material dispersion and can stay roughly constant or increase with the cavity length.} These findings establish bulk $n$-doped Tellurium as a viable platform for polarization-sensitive, tunable terahertz lasers and open a pathway toward exploiting BCD-induced gain in 3D materials for nonreciprocal and topological photonic applications.

\appendix
\section{\textcolor{black}{Lasing Threshold for Higher-Order Longitudinal Modes}}
\label{App: Higher-order modes}
\textcolor{black}{We examine the lasing threshold and frequency of oscillation of higher-order longitudinal modes by considering Eqs.~\eqref{eq:generalLasingCondition} and \eqref{eq:lasing_2} for $m>1$}. 

\begin{figure*}[tb!]
  \centering
  \includegraphics[scale=0.48]{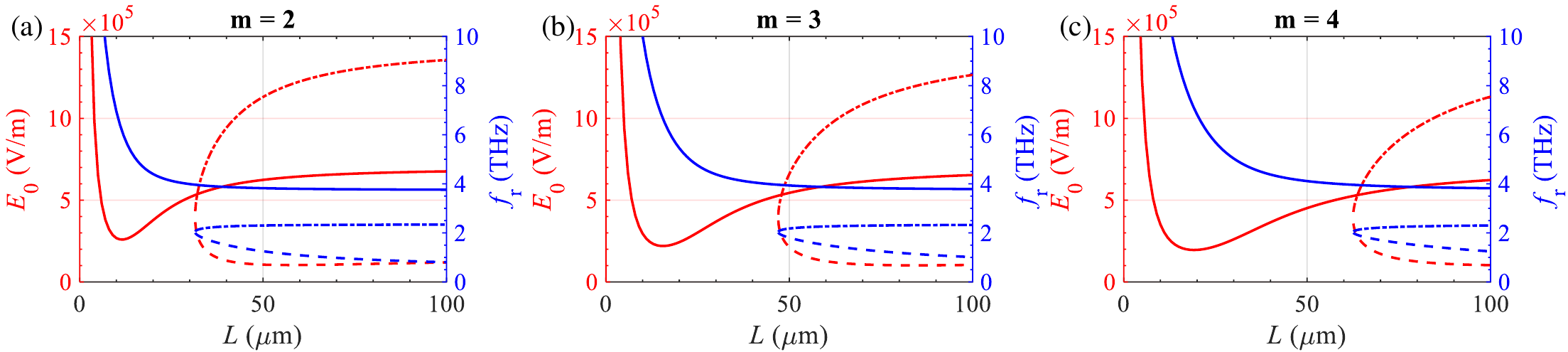} 
  \caption{\textcolor{black}{Case I. Biasing threshold for lasing, and corresponding frequency of oscillation for higher-order longitudinal modes of the cavity: (a) $m$ = 2; (b) $m$ = 3; (c) $m$ = 4.} }
  \label{fig:higherModes}
\end{figure*}

\textcolor{black}{For Case I, the solutions for $m=2,3,4$ are shown in Fig.~\ref{fig:higherModes}. Similar to Fig.~\ref{fig:CaseI_Lasing}, the solid curve in Fig.~\ref{fig:higherModes} corresponds to frequencies above the intrinsic resonance of Tellurium described by the Lorentz model in Eqs.~\eqref{eq:DielectricPermittivityLorentz1} and \eqref{eq:DielectricPermittivityLorentz2}. For this solid curve, at short cavity lengths, the oscillation frequency increases with the mode number $m$. However, at longer cavity lengths, the resonance frequency does not grow with the modal index $m$, and it approaches approximately $4\ \mathrm{THz}$, which is consistent with the behavior of $\alpha$ and $\beta$ in Case I. Furthermore, at long cavity lengths, we observe two solutions associated with the Lorentz resonance (dashed and dot-dashed curves). Interestingly, these lasing conditions require low biasing, and the frequency does not vary significantly by increasing $m$.}

\begin{figure*}[tb!]
  \centering
  \includegraphics[scale=0.48]{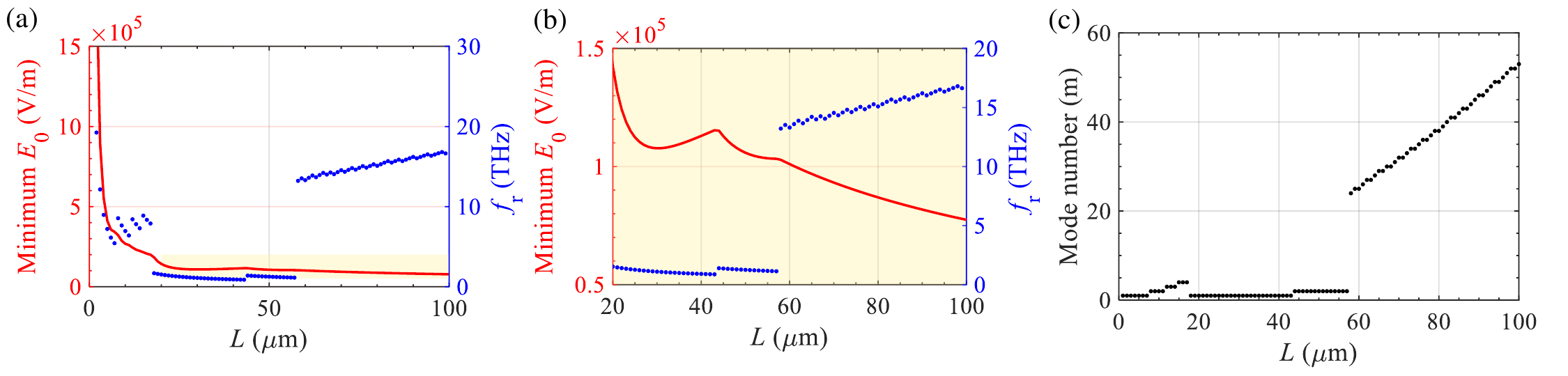} 
  \caption{\textcolor{black}{Case I. (a) Minimum bias among all modes with $m=1,2,..., 500$, required to initiate lasing, at various cavity lengths $L$, along with the corresponding oscillation frequency. (b) Magnified view of the boxed region in (a). (c) Mode number that yields the minimum required bias for lasing, varying cavity length.}} 
  \label{fig:minimumE0CaseI}
\end{figure*}

\begin{figure*}[tb!]
  \centering
  \includegraphics[scale=0.48]{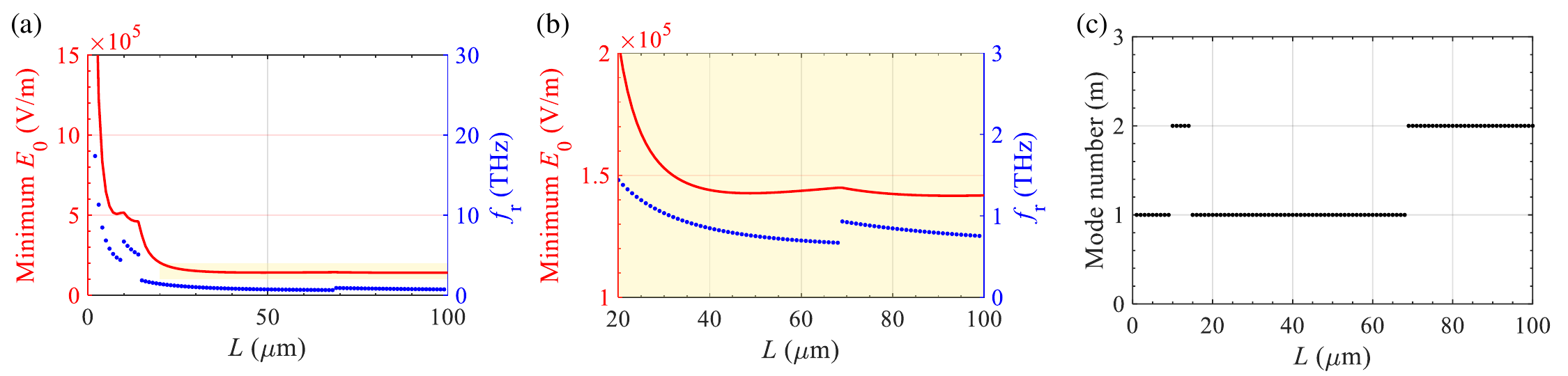} 
  \caption{\textcolor{black}{Same as Fig. \ref{fig:minimumE0CaseI}, but for Case II.} } 
  \label{fig:minimumE0CaseII}
\end{figure*}

\begin{figure*}[tb!]
  \centering
  \includegraphics[scale=0.48]{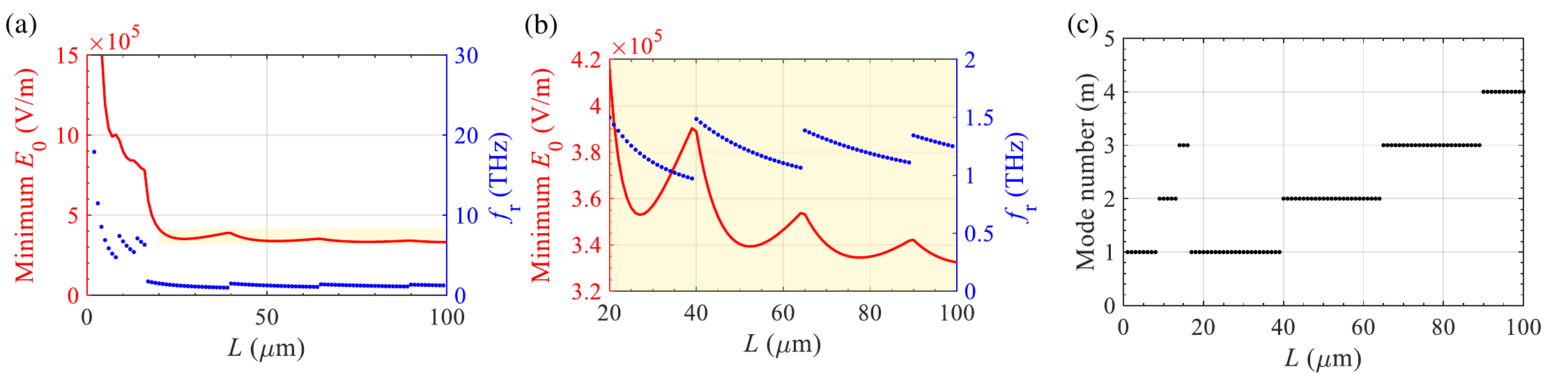} 
  \caption{\textcolor{black}{Same as Fig. \ref{fig:minimumE0CaseI}, but for Case III.} } 
  \label{fig:minimumE0CaseIII}
\end{figure*}

\textcolor{black}{
At each cavity length, we determine the minimum bias field $E_0$ required for a mode to start lasing. This threshold is plotted in Fig. \ref{fig:minimumE0CaseI} for Case I. The abrupt jumps occur when additional solution branches or higher-order modes with lower threshold values appear.
From the zoomed view in Fig.~\ref{fig:minimumE0CaseI}(b), we observe that for $L = 100\ \mu\mathrm{m}$, the minimum bias $E_0$ required to start lasing is $7.74\times10^4\ \mathrm{V/m}$, with the resonance frequency of 16.78 THz, and occurs for $m= 53$. The same computation is also carried out for Cases II and III, with the results presented in Figs. \ref{fig:minimumE0CaseII} and \ref{fig:minimumE0CaseIII}, respectively. In these two cases, the minimum bias happens at smaller mode numbers compared to Case I. For a cavity length of $L = 100\ \mu\mathrm{m}$, the minimum $E_0$ required to initiate lasing is $1.42\times10^5\; \mathrm{V/m}$ for Case II and $3.32\times10^5\;\mathrm{V/m}$ for Case III, corresponding to oscillation frequencies of 0.75 THz and 1.24 THz, respectively. This indicates that, by considering higher-order modes, lasing can be achieved at smaller bias values while maintaining THz oscillation frequencies in micrometer-scale cavities.}

\bibliography{References}
\end{document}